\documentclass[12pt]{article}

\input{tcilatex}

\begin{document}

\title{BTZ-like black holes in even dimensional Lovelock theories }
\author{Fabrizio Canfora$^{1}$, Alex Giacomini$^{2}$. \\
$^{1}${\small \textit{Centro de Estudios Cientificos (CECS), Casilla 1469
Valdivia, Chile.}}\\
$^{2}${\small \textit{Instituto de Fisica, Facultad de Ciencias, Universidad
Austral de Chile, Valdivia, Chile.}}\\
{\small e-mails: \textit{canfora@cecs.cl, alexgiacomini@uach.cl}}}
\maketitle

\begin{abstract}
In the present paper, a new class of black hole solutions is constructed in
even dimensional Lovelock Born-Infeld theory. These solutions are
interesting since, in some respects, they are closer to black hole solutions
of an odd dimensional Lovelock Chern-Simons theory than to the more usual
black hole solutions in even dimensions. This hybrid behavior arises when
non-Einstein base manifolds are considered. The entropies of these solutions
have been analyzed using Wald formalism. These metrics exhibit a quite
non-trivial behavior. Their entropies can change sign and can even be
identically zero depending on the geometry of the corresponding base
manifolds. Therefore, the request of thermodynamical stability constrains
the geometry of the non-Einstein base manifolds. It will be shown that some
of these solutions can support non-vanishing torsion. Eventually, the
possibility to define a sort of topological charge associated with torsion
will be discussed.
\end{abstract}

\section{Introduction}

The idea that space-time may have more than four dimensions goes back to
Kaluza and Klein in their attempt to find a geometrical unification of
gravity with electromagnetism \cite{kaluza-klein}. The idea of extra
dimensions is also very natural in the context of string theory (as a
classical textbook on the subject see e.g.\cite{strings}) and braneworld
scenarios \cite{randall-sundrum}. Thus, the question arises of which is the
most general classical theory of gravity in dimension higher than four. To
answer this question, one can use a gravitational action which is
constructed according to the same criteria as the action of gravity general
relativity in four dimensions: one requires that the action must be built
from curvature invariants which lead to second order differential equations
for the metric. Unlike the four-dimensional case (in which the only action
satisfying the previous conditions is the Einstein-Hilbert action plus a
cosmological term), in higher than four dimensions there are many more
possibilities: this class of theories is called Lovelock gravity \cite%
{Lovelock}. As it has been recently pointed out, Lovelock theories can
provide one with a quite intriguing phenomenological scenario for
compactification down to four dimensions \cite{CGTW} which is also able to
account for the smallness of the cosmological constant. For these reasons,
Lovelock gravities are worth to be further investigated.

A peculiar feature of Lovelock gravities is the richness of black hole
solutions they admit. Due to the non-trivial structure of the Lagrangian,
Lovelock black holes exhibit different types of behavior when compared to
the usual black hole thermodynamics of general relativity (see, for
instance, \cite{Gaston}). The focus of the present paper will be both on the
construction of some new black hole solutions in Lovelock theory and,
mainly, on the analysis of the interesting thermodynamical properties of
such solutions as well as on the possibility to provide one with "positive
entropy bounds" on the geometry of the base manifold.

An interesting class of Lovelock gravities corresponds to the case in which
the coupling constants of the theory are chosen in such a way that all the
possible $[(D-1)/2]$ maximally symmetric vacua degenerate. In even
dimensions this subclass of Lovelock theories is called Born-Infeld gravity.
In odd dimensions this request leads to an enhanced local symmetry group and
the theory becomes equivalent to a gauge theory with Chern-Simons action.
These two types of theories have very different physical properties. For
instance, the black solutions with positive constant curvature base
manifolds (Born-Infeld and Ba\~{n}ados-Teitelboim-Zanelli respectively) have
different lapse functions and thermodynamical behavior \cite{banados-zanelli}
, \cite{BHscan}. Moreover due to the enhanced symmetry of the Chern-Simons
theory, the equations of motion for the Black hole metric in odd dimensions
leave the base manifold completely arbitrary, while in even dimensions this
degeneracy does not occur. On the other hand, it is also known that in
Lovelock gravity the base manifold needs to be neither constant curvature
nor Einstein (\cite{Dotti-Oliva}, \cite{Dotti-Oliva2} \cite{Dotti-Oliva3}, 
\cite{BCharm},\ \cite{Hideki}). However, in the cases of Chern-Simons and
Born-Infeld Lovelock theories black hole solutions with non-Einstein base
manifold have not received a lot of attention.

The main goal of this paper is to explore the physical properties of black
hole solutions in Born-Infeld gravity by allowing the base manifold to be
non-Einstein. The most simple case of a non-Einstein manifold is a product
of several constant curvature base manifolds. Interestingly enough, in this
case the lapse function has not the usual even-dimensional Born-Infeld form
as in \cite{banados-zanelli}, \cite{BHscan} but the one of a Ba\~{n}%
ados-Teitelboim-Zanelli (\textbf{BTZ}) black hole in odd dimensions. These
solutions are therefore in some sense a hybrid between an odd and even
dimensional black hole as they have many features in common with
odd-dimensional solutions but lack of the high degeneracy of the
Chern-Simons case. If one chooses this BTZ-like lapse function, the
equations of motion of the theory give a constraint on the geometry of the
base manifold which forces it to be non-Einstein. If the base manifold is a
product of constant curvature spaces, the equation of motion gives a
relation between the curvature radii of the factors. In the case of a
constant curvature base manifold, it is always possible to "rescale" the
radial coordinate is such a way that the curvature radius can be normalized
to $\pm 1$ or zero. In the case of a base manifold which is the product of
constant curvature spaces, it is possible to rescale to $\pm 1$ only one of
the curvature radii of the constant curvature factors. This introduces a new
length scale which affects the thermodynamical behavior of the black holes.
In some cases, the entropy becomes a non-positive definite function of the
remaining curvature radius. The fact that in Lovelock gravities black holes
can have negative entropy was firstly discussed in \cite{cvetic}. In this
reference, the authors showed that the arising of negative entropies can be
attributed to the freedom in choosing the coupling constants in the Lovelock
gravities. However, in the present case, this happens when the Lovelock
coupling constants are fixed up to an overall factor in the action in such a
way that all the possible $[(D-1)/2]$ maximally symmetric vacua degenerate
(note that this choice of the coupling constants does not belong to the
cases discussed in \cite{cvetic}). The feature which allows the appearance
of negative entropies in the present case is the "non-Einstein nature" of
the base manifold. Since, as it has been already noticed in this context 
\cite{cvetic} and as it is well known in statistical mechanics (the example
of mean field spin glass is very clear \cite{MPV} in this respect), negative
entropy is a strong signal of instability, the present results provide one
with a thermodynamical bound on the possible geometries of non-Einstein base
manifolds. \newline

In this paper, six, eight and ten dimensional examples will be considered in
details as they catch all the fundamental features of the higher dimensional
cases.\newline
In six dimensions, within the class of base manifolds which are product of
constant curvature spaces, the base manifold must have the form $M_{2}\times
N_{2}$ where $M_{2}$ and $N_{2}$ are two-dimensional non-vanishing constant
curvature manifolds. In higher than six dimensions, there are more ways to
construct base manifolds as product of constant curvature spaces: these new
possibilities give rise to different thermodynamical behaviors. In ten
dimensions, we will focus on the class of base manifolds with the form $%
M_{4}\times N_{4}$ ($M_{4}$ and $N_{4}$ being four-dimensional non-vanishing
constant curvature manifolds): in this case, the solution presents different
branches which can be compared using thermodynamical arguments. In eight
dimensions, a case of special interest is the one in which the
six-dimensional base manifold is the product of two non-vanishing constant
curvature three dimensional manifolds, $M_{3}$ and $N_{3}$: the entropy of
the corresponding black hole solution identically vanishes. Moreover, with
this base manifold, the vacuum solution allows the presence of non-trivial
torsion. This is a quite non-trivial result: unless the ansatz for the
torsion is chosen carefully as in \cite{canfo}, one gets an over-constrained
system of equations which "kills" the torsional degrees of freedom. In odd
dimensions, in the case of Chern-Simons (\textbf{CS}) Lovelock gravities,
the further consistency conditions on the torsion disappear because of the
enhanced gauge symmetry of the theory. A black hole with torsion in this
case has been constructed in \cite{aros}. In even dimensions, the
Born-Infeld Lovelock gravity is somehow similar to the CS case since the
Lovelock coupling constants are tuned in such a way that there exists only
one maximally symmetric vacuum. On the other hand, in the Born-Infeld case,
no enhanced gauge symmetry appear and the consistency conditions on the
torsion \textit{are not} identically satisfied (anyway, as it has been shown
in \cite{TROZAN} the Born-Infeld Lovelock gravity is the one where the
torsion acquires the maximal number of degrees of freedom). When the torsion
lives on a three-dimensional sub-manifold, it can be proportional to the
three-dimensional completely antisymmetric tensor whose properties imply
that most of the consistency conditions are automatically satisfied. This
ansatz, proposed for the first time in \cite{canfo}, allowed to find the
first exact vacuum solutions with non-vanishing torsion in a
five-dimensional Lovelock theory without enhanced gauge symmetry\footnote{%
An amusing feature of this solution is that it can be interpreted as the
purely gravitational analogue of the Bertotti-Robinson space-time.} \cite%
{CGW}. For a five-dimensional black hole metric, such ansatz for torsion
turns out to be consistent in the CS case \cite{CGT}. So it is natural to
ask if there exist black hole solutions with torsion in the cases in which
no enhanced symmetry is present. An interesting possibility explored in \cite%
{CG} is to consider a eight-dimensional metric which represents a
five-dimensional black hole with three compact extra dimensions: in this
case the metric supports a non-vanishing torsion flux. Instead here we will
construct a genuine eight-dimensional black hole with non-zero torsion along
the $M_{3}\times N_{3}$ base manifold. \ Eventually, the present
construction strongly suggests that it should be also possible to associate
to this torsion a conserved charge closer to a monopole charge than to a
Noether charge.\newline

The structure of the paper will be the following: In the second section, the
black hole ansatz will be discussed. In the third and fourth sections, the
six and eight dimensional cases will be discussed and the corresponding
thermodynamical features will be analyzed using the Wald formula \cite%
{waldentropy} \cite{waldentropy2}. It will be also discussed that, in the
eight dimensional case, torsion can be introduced in a very natural way. In
the fifth section, the possibility to define a topological charge associated
to the torsion is analyzed. In the sixth section the ten dimensional case is
considered. In the last section, some conclusions are drawn.

\section{The Black Hole ansatz}

A newsworthy class of Lovelock theories in $D$ dimensions is the one for
which all the possible $[(D-1)/2]$ maximally symmetric vacua degenerate to
one vacuum. This corresponds in odd dimensions to a Chern-Simons (CS) theory
and in even dimensions to a Born-Infeld (BI) theory. The static black hole
solutions with a spherical base manifold are given by 
\begin{equation}
ds^{2}=-f^{2}(r)dt^{2}+\frac{dr^{2}}{f^{2}(r)}+r^{2}d\Omega ^{2}  \label{cs}
\end{equation}%
Where the function $f$ is given by 
\begin{equation}
f^{2}(r)=r^{2}/l^{2}-\mu  \label{btzlapse}
\end{equation}%
in the odd dimensional case. It is important to stress that in the CS case
the equations of motion leave the base manifold arbitrary\footnote{%
However, in five dimensions, at least in the case of zero-mass black hole,
the degeneracy can be removed by requiring that the solution preserves half
of the supersymmetries of the corresponding Chern-Simons supergravity \cite%
{CGT}: this can be only achieved by introducing torsion in a suitable way.}.
In the even dimensional case $f$ is given by 
\begin{equation}
f^{2}(r)=r^{2}/l^{2}+1-\frac{2M}{r}^{D/2-1}\ \ ,  \label{BIlapse}
\end{equation}%
where, unlike the case of CS Lovelock gravities, the base manifold is
determined by the equations of motion to be (in the case of the lapse
function in Eq. (\ref{BIlapse})) of constant curvature. Thus, in the case of
BI Lovelock gravities, static spherically symmetric black holes whose base
manifolds have positive constant curvature cannot have the lapse function in
Eq. (\ref{btzlapse}). On the other hand, many of the nice features of the
thermodynamics of CS Lovelock black holes (see for a nice review \cite%
{Gaston}) are closely connected to the lapse function in Eq. (\ref{btzlapse}%
). Therefore, it is very interesting to try to "push" the BI Lovelock theory
in even dimensions to see if it is possible to construct some black hole
solutions close enough to the BTZ-like black hole solutions of CS Lovelock
gravities.

Indeed, this goal can be achieved provided one allows the base manifold to
be non-Einstein. The most simple example of a non-Einstein manifold is a
product of several constant curvature spaces. Interestingly enough, it will
be shown here that in this case the lapse function can have the BTZ-like
form in Eq. (\ref{btzlapse}).

The gravitational action of the even dimensional Born-Infeld Lovelock theory
in $D=2n$ dimensions can be written in the first order formalism as 
\begin{equation}
L_{2n}=\kappa \int \epsilon _{a_{1}\ldots }a_{2n}\overline{R}%
^{a_{1}a_{2}}\ldots \overline{R}^{a_{2n-1\;2n}}  \label{bi1}
\end{equation}%
where we have defined 
\[
\overline{R}^{a_{i}a_{j}}=R^{a_{i}a_{j}}+\frac{e^{a_{i}}e^{a_{j}}}{l^{2}}\ , 
\]%
$R^{a_{i}a_{j}}$ is the curvature two form 
\[
R^{a_{i}a_{j}}=d\omega ^{a_{i}a_{j}}+\omega _{\ \ a_{k}}^{a_{i}}\omega
^{a_{k}a_{j}}\ \ , 
\]%
$\omega ^{a_{i}a_{j}}$ is the spin connection and $e^{a_{i}}$ is the
vielbein and the overall coupling constant $\kappa $ will be discussed in
the next sections. In literature it is customary to call the coupling
constant of the n-th order Lovelock term $c_n$.The Born-Infeld tuning
between the coupling constants $c_n$ then, according to (\ref{bi1}), is
found by just applying Newton's binomial formula. The equations of motion
obtained by varying the action with respect to the vielbein and the spin
connection read respectively 
\begin{equation}
E_{a_{i}}=\epsilon _{a_{1}\ldots a_{2n}}\overline{R}^{a_{2}a_{3}}\ldots 
\overline{R}^{a_{2n-2}\,a_{2n-1}}e^{a_{2n}}=0\ .  \label{eombi}
\end{equation}%
\begin{equation}
E_{a_{i}a_{j}}=\epsilon _{a_{1}\ldots a_{2n}}T^{a_{1}}\overline{R}%
^{a_{2}a_{3}}\ldots \overline{R}^{a_{2n-4}\,a_{2n-3}}e^{a_{2n-2}}=0\ .
\label{torsion}
\end{equation}

In this section it will be considered the torsion free case so that the
following identity holds:%
\[
T^{a_{i}}=De^{a_{i}}=de^{a_{i}}+\omega _{\ \ a_{k}}^{a_{i}}e^{a_{k}}=0 
\]%
where $T^{a_{i}}$ is the torsion field. The ansatz for the black hole metric
is 
\begin{equation}
ds^{2}=-f^{2}dt^{2}+\frac{dr^{2}}{f^{2}}+d\Sigma ^{2}  \label{bhansatz}
\end{equation}%
where $\Sigma $ is a $2n-2$ manifold which is the product of constant
curvature manifolds. The lapse function is of the form 
\begin{equation}
f^{2}=r^{2}/l^{2}-\mu \ \ ,  \label{klapse}
\end{equation}%
where $\mu $ is a mass parameter since a non-trivial $\mu $ allows the
presence of a black hole horizon. The vielbein are of the form 
\begin{eqnarray*}
e^{0} &=&fdt\ ;\ \ e^{1}=\frac{dr}{f}\ \ , \\
e^{i} &=&r\tilde{e}^{i}\ \ \ ,
\end{eqnarray*}%
where $\tilde{e}^{i}$ are the intrinsic vielbein of the base manifold.

With the ansatz in Eqs. (\ref{bhansatz}) and (\ref{klapse}) the nontrivial
equations of motion are 
\begin{eqnarray}
E_{0} &=&\epsilon _{01i_{1}\ldots i_{2n-2}}\overline{R}^{i_{1}i_{2}}\ldots 
\overline{R}^{i_{2n-3}i_{2n-2}}e^{1}=0\ \ ,  \label{eq1} \\
E_{1} &=&\epsilon _{10i_{1}\ldots i_{2n-2}}\overline{R}^{i_{1}i_{2}}\ldots 
\overline{R}^{i_{2n-3}i_{2n-2}}e^{0}=0\ \ .  \label{eq2}
\end{eqnarray}%
In the case of a constant curvature base manifold, the above equations of
motion would have forced the mass parameter to be $\mu =-1$ preventing the
appearance of a horizon. On the other hand, as it will be shown in the next
section, when the base manifold is the product of two (or more) constant
curvature manifolds, a black hole horizon can appear.

\section{Six dimensional black holes}

The simplest example in six dimensions is the case in which the base
manifold has the form $M_{(2)}\times N_{(2)}$ (where $M_{(2)}$ and $N_{(2)}$
are manifolds of constant curvatures with curvatures $\gamma $ and $\eta $
respectively). The metric reads then 
\begin{equation}
ds^{2}=-f^{2}(r)dt^{2}+\frac{dr^{2}}{f^{2}(r)}+r^{2}(dM_{2}^{2}+dN_{2}^{2})\
\ ,  \label{6dcase}
\end{equation}%
where $dM_{2}^{2}$\ and $dN_{2}^{2}$\ are the intrinsic metrics of $M_{2}$\
and $N_{2}$. The indices along the manifold $M_{2}$ will be denoted as $i$, $%
j$, $k$,... while the indices along $N_{2}$ will be denoted as $a$, $b$, $c$%
... With this notation, the only nonzero curvatures are 
\[
\overline{R}^{ij}=\frac{\gamma +\mu }{r^{2}}e^{i}e^{j}\ ,\ \ \ \overline{R}%
^{ab}=\frac{\eta +\mu }{r^{2}}e^{a}e^{b}\ ,\ \ \ \overline{R}^{ia}=\frac{\mu 
}{r^{2}}e^{i}e^{a}\ ,
\]%
so that the equations of motion Eqs. (\ref{eq1}) and (\ref{eq2}) (in the
present case the two equations of motion are not independent, we will also
assume that $\mu \neq 0$) read%
\begin{eqnarray}
(\mu +\gamma )(\mu +\eta )+2\mu ^{2} &=&0\;\Leftrightarrow   \label{tulin0}
\\
(1+\frac{\gamma }{\mu })(1+\frac{\eta }{\mu })+2 &=&0\;,  \label{tulin1}
\end{eqnarray}
thus, the two relevant parameters are%
\[
\gamma ^{\prime }=\frac{\gamma }{\mu }\ \ \ \ and\ \ \ \ \eta ^{\prime }=%
\frac{\eta }{\mu }.
\]
In the present case it is convenient to rescale the mass parameter $\mu $ in
the lapse function to $1$ since this allows to keep track easily of the
curvature scales. Obviously, by fixing the mass parameter $\mu $, \textit{%
one is not fixing the mass} as the physical mass is also proportional to the
volume of the base manifold which depends on the curvature scales $\gamma $
and $\eta $. Note also that the case $\mu =0$ does not represent a black
hole. With this normalization of the mass parameter, the equation of motion
reads 
\begin{equation}
(1+\gamma )(1+\eta )+2=0\;\Rightarrow   \label{tulin3}
\end{equation}%
\begin{equation}
\gamma =-\frac{\left( 3+\eta \right) }{1+\eta }\ \ ,\;\;\eta \neq -1\
,\;\;\;\gamma \neq -1\ .  \label{firsteom}
\end{equation}%
In this six dimensional case (in which BI gravity is a particular case of
Einstein-Gauss-Bonnet gravity), the expression in Eq. (\ref{firsteom}) is
the first concrete solution of the constraint proposed in  \cite{Dotti-Oliva}%
, \cite{Dotti-Oliva2} \cite{Dotti-Oliva3} for Einstein-Gauss-Bonnet gravity
in arbitrary even dimensions in the case of degenerate vacua. Using the
above expression for $\gamma $, one can easily check that there is no choice
of $\eta $ for which $\gamma =\eta $ in such a way that a BTZ lapse function
does not allow Einstein base manifolds. As the radial coordinate has been
already used to normalize the mass parameter, the curvature radius $\eta $
is indeed an integration constant. Looking at the equation (\ref{firsteom})
it is easy to realize that the possible base manifolds are $H_{2}\times H_{2}
$ and $H_{2}\times S_{2}$ while the case of $S_{2}\times S_{2}$ is not
realized since $\gamma $ and $\eta $\ cannot be both positive (here, $S_{2}$
represents a two-dimensional compact positive constant curvature space while 
$H_{2}$ represents a two-dimensional compact negative constant curvature
space).\textbf{\ }

It is worth pointing out that for base manifolds which are the product of
constant curvature spaces\textit{\ in which there is only one factor with
non-vanishing curvature} one would have the curvature radius and the mass
completely fixed by the equations of motion. Such isolated points in the
solution space are of less physical interest as in such a case it is not
obvious how to achieve a satisfactory thermodynamical description.
Eventually, a base manifolds which is a product of vanishing curvature
spaces is not compatible with the ansatz in Eqs. (\ref{bhansatz}) and (\ref%
{klapse}).\textbf{\ }

As far as the class of metrics which we are analyzing is concerned, the
solutions must have a base manifold where at least two factors have
non-trivial curvature. This means that in six dimensions the base manifolds $%
H_{2}\times H_{2}$ or $H_{2}\times S_{2}$ are the only possibilities. The
intrinsic volume of the corresponding base manifold (which is important in
order to compute the mass) reads 
\begin{equation}
Vol\left( M_{2}\times N_{2}\right) =\xi \frac{(1+\eta )^{2}}{\eta
^{2}(3+\eta )^{2}}  \label{volume}
\end{equation}%
where $\xi $\ is a positive combinatorial factor which will not play any
role in the following. The entropy of this solution can be computed using
the Wald formula \cite{waldentropy}. For Lovelock theories, there is a
simple way to apply the Wald formula \cite{waldentropy2}: the $n$-th
Lovelock term "$c_{n}R^{n}$" in $D$ dimensions (which has $n$ factors of the
curvature two form) in the action ,%
\[
c_{n}R^{n}\approx c_{n}\epsilon
_{a_{1}..a_{D}}R^{a_{1}a_{2}}..R^{a_{2n-1}a_{2n}}..e^{a_{D}}\ ,
\]%
($c_{n}$ being the coupling constant of the $n-$th term in the Lovelock
series) contributes to the entropy with a term $S^{(n)}$ of the form%
\begin{equation}
S^{(n)}=nc_{n}\int_{H}\epsilon
_{a_{1}..a_{D-2}}R^{a_{1}a_{2}}..R^{a_{2n-3}a_{2n-2}}..e^{a_{D-2}}
\label{entrop0w}
\end{equation}%
where the integral is evaluated on the horizon $H.$ Thus, using the BI
tuning for the coupling constants, the contribution $S^{(n)}$ depends on the 
$(n-1)$-th Lovelock term evaluated on the horizon $H$. Then, the total Wald
entropy reads 
\begin{equation}
S=\sum S^{(n)}\ .  \label{entropyw}
\end{equation}%
In six dimensions the highest nontrivial Lovelock term is the quadratic
Gauss-Bonnet term while, as usual, the contribution of the Einstein-Hilbert
term is proportional to the volume of the base manifold: thus, the total
entropy in Eq. (\ref{entropyw}) reads 
\begin{equation}
S_{6D}=S_{6D}(\eta )=\kappa c_{1}r_{+}^{4}Vol\left( M_{2}\times N_{2}\right) 
\left[ \frac{2(\eta ^{2}-3)}{1+\eta }+6\right] \ \ ,  \label{ent6}
\end{equation}%
where $r_{+}$ is the radius of the black hole horizon (which in our units is 
$r_{+}=\left\vert l\right\vert $). Inserting the explicit expression for the
intrinsic volume of the base manifold and (\ref{tulin1}) the entropy becomes 
\begin{equation}
S_{6D}=2\kappa \xi r_{+}^{4}c_{1}\frac{\eta +1}{(\eta +3)\eta }
\label{entro6d}
\end{equation}%
It is worth pointing out that the above expression for the entropy in this
six-dimensional case, presents a quite interesting and novel feature. The
entropy has a zero in $\eta =-1$ where it changes sign. When the overall
coupling constant $\kappa $ of the six-dimensional BI-Lovelock action\ is
chosen to be positive (as it is usual \cite{banados-zanelli}), the entropy
in Eq. (\ref{entro6d}) is negative for $\eta <-3$ and $-1<\eta <0$. These
ranges for the integration constant correspond to a base manifold of the
form $H_{2}\times H_{2}$. In the present case, choosing a negative $\kappa $
would not help to avoid the arising of a negative entropy\footnote{%
In classical thermodynamics the entropy is defined up to an additive
constant. Therefore, one may wonder whether a suitable additive constant can
be found in order to avoid the appearence of negative entropy solutions. On
the other hand, as it can be seen in Eq. (\ref{entro6d}), in the present
cases the entropy can even approach to $-\infty $. Thus, it seems that the
appearence of negative entropy solutions cannot be avoided by adding a
suitable constant to the (Wald) entropy.}. The appearance of black hole
solutions with negative entropy in higher order theories has been already
pointed out in \cite{cvetic}: as known results in statistical mechanics
suggest (see, for instance, the classic review on spin glass \cite{MPV}), a
solution with negative entropy should decay into a more stable solution. In 
\cite{cvetic}, the origin of the negativity of the entropy was the
possibility to consider cases in which some of the coupling constants of the
theory can be arbitrary negative. In the present example, the coupling
constants are the ones characterizing the BI Lovelock Lagrangian and, in
particular, this choice of the coupling constants appears to be outside the
range of \cite{cvetic}. Therefore, in the present example the non-Einstein
nature of the base manifold (and in particular the appearance of a further
curvature scale) is behind the sign change of the entropy. Hence, the above
thermodynamical argument provides the geometry of the base manifold with a
stability constraint. A necessary condition in order for the black hole in
Eqs. (\ref{6dcase}), (\ref{klapse}) and (\ref{firsteom}) to be
thermodynamically stable is that when $\kappa >0$ the curvature $\eta $ of
the factor $N_{(2)}$ of the base manifold has to satisfy the following 
\textit{positive entropy bound}%
\[
\eta >0\ \ \ \ \ or\ \ \ \ \ 3<\eta <-1
\]%
This implies that the base manifold must be of the form $S_{2}\times H_{2}$
(the base manifold $S_{2}\times S_{2}$ was already excluded by the equations
of motion). If $\kappa <0$, one has to consider the complement of the above
conditions. Eventually, the mass of the black hole can be computed using the
first law of black hole thermodynamics which in the static case takes the
usual form \cite{waldentropy2} $\delta U=T\delta S$ where the temperature $T$
is given as usual in terms of the derivative of the lapse function at the
horizon $T=\frac{1}{4\pi }f^{\prime }$. As it is natural to expect, it is
indeed proportional to the intrinsic volume of the base manifold. In any
case, in the following analysis, it will be enough to study the behavior of
the Wald entropy.

\section{Eight dimensional black hole}

In eight dimensions more possibilities for the choice for the base manifolds
appear. Also in this case, it is convenient to normalize to one the mass
parameter $\mu =1$ since the mass parameter will be assumed to be
non-vanishing. The most interesting case is $M_{3}\times N_{3}$ where the
two factors are three dimensional constant curvature manifold with
(non-vanishing) curvatures $\gamma $ and $\eta $. The reason is that, with
this choice, the solution may also support a non-trivial torsion flux and
exhibits a very peculiar thermodynamics as it will be explained in the next
section. The metric reads 
\begin{equation}
ds^{2}=-f^{2}(r)dt^{2}+\frac{dr^{2}}{f^{2}(r)}+r^{2}(dM_{3}^{2}+dN_{3}^{2})\
\ ,  \label{8dcase}
\end{equation}%
where the lapse function is as in Eq. (\ref{klapse}). The equations of
motion with the useful normalization $\mu =1$ gives 
\begin{equation}
3(\gamma +1)(\eta +1)+2=0\ \ \Rightarrow \ \gamma =-\left( \frac{3\eta +5}{%
3(\eta +1)}\right) \ .  \label{8dequation1}
\end{equation}%
Using the above expression for $\gamma $, one can easily check that also in
this case there is no choice of $\eta $ which allows $\gamma =\eta $. In
other words, if the lapse function has the BTZ form then the base manifold
cannot be Einstein. The intrinsic volume of the base manifold is 
\[
Vol(M_{3}\times N_{3})=\xi \frac{1}{\left\vert \gamma \eta \right\vert ^{3}}%
=\xi |\frac{27(\eta +1)^{3}}{\eta ^{3}(3\eta +5)^{3}}|
\]%
where $\xi $\ is a positive combinatorial factor which will not be important
for the following analysis. Recalling that in eight dimensions the highest
non-trivial Lovelock term is cubic in the curvature, Eqs. (\ref{entrop0w}), (%
\ref{entropyw}), leads to the following expression for the Wald entropy%
\begin{eqnarray*}
S &=&Vol(M_{3}\times N_{3})\left( r_{+}\right) ^{6}\left[ 3\frac{c_{3}}{l^{4}%
}\left( 4\gamma \eta \right) +\right.  \\
&&\left. +2\frac{c_{2}}{l^{2}}\left( 4\gamma +4\eta \right) +20c_{1}\right]
\ .
\end{eqnarray*}
Using the BI tuning of the Lovelock coupling constants in eight dimensions,%
\[
\frac{c_{3}}{c_{2}}=\frac{2l^{2}}{3}\ ,\ \ \ \frac{c_{1}}{c_{2}}=\frac{2}{%
3l^{2}}\ ,
\]
as well as Eq. (\ref{8dequation1}), it is easy to see that the entropy of
this black hole identically vanishes for any value of the integration
constant $\eta $ 
\[
S\equiv 0\,.
\]%
It is worth to point out that the temperature of this black hole remains
finite. Cases of black holes with zero entropy and finite temperature are
known in literature in the context of compactified Lovelock black holes as
well as in the context of Lifshitz black holes \cite{cai} \cite{cai2}%
\footnote{%
Other papers somehow related to this subject are \cite{cai3}.}. In these
cases the vanishing of the entropy is related to a specific tuning of the
coupling constants rather than to the topology of the base manifold. For the
class of black hole solutions studied in the present paper, the existence of
a solution with zero entropy is due to the special topology (in a sense that
will be explained in the next section) of the base manifold $M_{3}\times
N_{3}$.

\subsection{$M_4\times N_2$ base manifold}

A black hole solution with zero entropy is a quite peculiar feature of the $%
M_{3}\times N_{3}$ base manifold. This can be seen as follows. In eight
dimensions there is, for instance, the possibility to choose a less
symmetric base manifold such as $M_{4}\times N_{2}$. Let the curvature
radius of the $M_{4}$ factor be $\gamma $ and the one of $N_{2}$ factor be $%
\eta $. The equations of motion imply the relation 
\begin{equation}
\eta =-\frac{5+\gamma }{1+\gamma }
\end{equation}%
The entropy of the corresponding black hole reads 
\begin{equation}
S=3\kappa \xi r_{+}^{6}c_{1}\frac{(1+\gamma )^{2}}{\gamma ^{4}(5+\gamma )^{2}%
}\left[ \gamma ^{2}+6\gamma -\frac{5+\gamma }{1+\gamma }(2\gamma +1)+5\right]
\end{equation}%
($\xi $ being a positive combinatorial factor which plays no role in the
present analysis) which is negative for $-5<\gamma <-1$. This range of
parameters corresponds to a base manifold of the form $H_{4}\times S_{2}$.
It is worth to point out that the entropy corresponding to the base manifold 
$S_{4}\times H_{2}$ is positive. Thus, in particular, the entropy in this
case is not identically zero. Analogous computations for a more general
choice of base manifolds as product of constant curvature spaces reveal that
an identically vanishing entropy is a peculiarity of the $M_{3}\times N_{3}$%
\ base manifold.

\section{non-trivial torsion}

Besides the identically vanishing entropy, the eight dimensional black hole
with base manifold $M_3\times N_3$ has another very interesting feature:
namely it supports a non trivial torsion flux according to the prescription
of \cite{canfo}. In particular, this appears to be the first black hole
solution with non-vanishing torsion\footnote{%
It is worth noting that, in \cite{CG}, it has been constructed a
compactified black hole solution with torsion in eight dimensions: namely an
eight-dimensional metric which represent a five-dimensional black hole with
three compactified extra dimensions. While the black hole solution with
torsion found in \cite{CGT} corresponds to a Chern-Simons Lovelock theory
with, therefore, enhanced gauge symmetry.} in the case of a Lovelock theory
without enhanced gauge symmetry. The following ansatz for the torsion 
\begin{equation}
T^{i}=F_{1}(r)\epsilon ^{ijk}e_{k}e_{k}\ ,\ \ \ T^{a}=F_{2}(r)\epsilon
^{abc}e_{b}e_{c}\;,  \label{ansatzt1}
\end{equation}%
and the contorsion tensor 
\begin{equation}
K^{ij}=-F_{1}(r)\epsilon ^{ijk}e_{k}\;\;\;;\;\;\;K^{ab}=-F_{2}(r)\epsilon
^{abc}e_{c}  \label{contorsion}
\end{equation}%
with 
\begin{equation}
F_{1}=\frac{\delta _{1}}{r}\;\;\;;\;\;\;F_{2}=\frac{\delta _{2}}{r}
\label{ansatz1.5}
\end{equation}%
is very natural since the only modifications of the curvature two form
(which now receives contributions also from the contorsion tensor) are 
\begin{equation}
R^{1i}=\hat{R}^{1i}-\frac{f}{r}T^{i}\;\;\;;\;\;\;R^{ij}=\hat{R}^{ij}-\left( 
\frac{\delta _{1}}{r}\right) ^{2}e^{i}e^{j}  \label{modifiedcurvature}
\end{equation}%
\begin{equation}
R^{1a}=\hat{R}^{1a}-\frac{f}{r}T^{a}\;\;\;;\;\;\;R^{ab}=\hat{R}^{ab}-\left( 
\frac{\delta _{2}}{r}\right) ^{2}e^{a}e^{b}\ \ \Rightarrow
\label{modifiedcurvature2}
\end{equation}%
\begin{eqnarray}
R^{ij} &=&\frac{\tilde{\gamma}-f^{2}}{r^{2}}e^{i}e^{j}\;,\;\;\;R^{ab}=\frac{%
\tilde{\eta}-f^{2}}{r^{2}}e^{a}e^{b}\ \ ,  \label{modified3} \\
\tilde{\gamma} &=&\gamma -\left( \delta _{1}\right) ^{2}\ \ ,\ \ \ \tilde{%
\eta}=\eta -\left( \delta _{2}\right) ^{2}\ \ \ ,  \label{modified4}
\end{eqnarray}%
where $\gamma $ and $\eta $\ denote the Riemannian torsion-free parts of the
scalar curvatures of the two factors of the base manifold, $\hat{R}^{ab}$, $%
\hat{R}^{ij}$, $\hat{R}^{1i}$, $\hat{R}^{1a}$,... denote the Riemannian
torsion-free parts of the curvature while $R^{ab}$, $R^{ij}$, $R^{1i}$, $%
R^{1a}$,... denote the total curvature. It is easy to see that with this
choice of the torsion, the equations of motion obtained by varying the spin
connection for the torsion (\ref{torsion}) are identically satisfied so that 
$\delta _{1}$ and $\delta _{2}$ are integration constants. The equations of
motion obtained by varying the action with respect to the vielbein (\ref%
{eombi}) reduce once again to Eq. (\ref{8dequation1}) \textit{but with the
replacement} 
\begin{equation}
\gamma \rightarrow \tilde{\gamma}=\gamma -\left( \delta _{1}\right) ^{2}\ \
,\ \ \ \eta \rightarrow \tilde{\eta}=\eta -\left( \delta _{2}\right) ^{2}\ .
\label{torsionreplacement}
\end{equation}%
The technical reason for this lies in the nice identities satisfied by a
torsion of the form in Eq. (\ref{ansatzt1}): see, for instance, \cite{canfo}
and \cite{CG}. Interestingly enough, as it will be explained in more details
in the next section, the geodesics of the metric in Eq. (\ref{8dcase}) with
no-vanishing torsion fluxes as in Eqs. (\ref{ansatzt1}) and (\ref{ansatz1.5}%
) along the two factors $M_{(3)}$ and $N_{(3)}$ "almost" do not feel the
torsion. Namely, the torsion enters the metric only in Eq. (\ref{8dequation1}%
) through the replacements in Eq. (\ref{torsionreplacement}) which affects
the relation between $\gamma $ and $\eta $:%
\[
\tilde{\gamma}=-\left( \frac{3\tilde{\eta}+5}{3(\tilde{\eta}+1)}\right) \ \
\Leftrightarrow \ \ \gamma =\left( \delta _{1}\right) ^{2}-\left( \frac{%
3\left( \eta -\left( \delta _{2}\right) ^{2}\right) +5}{3(\eta -\left(
\delta _{2}\right) ^{2}+1)}\right) . 
\]%
On the other hand, being the torsion fully anti-symmetric, the corresponding
contorsion tensor does not affect directly the geodesic equation: see, for
instance, \cite{CG}. However, the dynamics of Fermions is affected by the
presence of torsion: the contorsion tensor directly enters the Dirac
equation modifying the spin connection in a way very similar to a
"spin-spin" coupling in which the contorsion plays the role of "spin" of the
gravitational background. At present, a formulation of Wald entropy in the
presence of non-vanishing torsion fluxes is still unavailable but it can be
argued that torsion could also have, for suitable choices of the integration
constants $\delta _{1}$\ and $\delta _{2}$, positive effects on the
stability of the solution (as it was first shown in \cite{CGT}). A more
detailed discussion on this point is in the following section.

\subsection{topological charge?}

It is very interesting to discuss how the presence of torsion may affect the
definition of charges and, in particular, the possibility to construct some
sort of topological charge: this issue is closely connected to the question
of stability. In general, one should expect that the torsion affects the
definition of conserved charges in (at least) two ways.

Firstly, when defining the conserved charges in Lovelock gravities using
asymptotic integrals and counterterms \cite{regge-teitelboim} \cite%
{troncoso-olea} or using the modified version of the Komar integrals \cite%
{kastor} one always uses the absence of torsion to derive some useful
identities (in a recent paper \cite{gravanis} the author has shown how to
include possible torsion contributions). For this reason, in the presence of
a non-trivial torsion, one should expect that the usual charges get some
extra contributions.

Secondly and more interestingly from the point of view of the stability
analysis, one should also expect that the presence of torsion could give
rise to some sort of \textit{topological charge}.

To clarify this point, let us briefly recall some well known features of
Yang-Mills-Higgs theory. In this case, as it is well known, the (non-Abelian
generalization of the) electric charges can be constructed, for instance,
using the Noether method. In order to perform such procedure, one needs to
use both the symmetries of the action and the equations of motion. On the
other hand, in non-Abelian Yang-Mills-Higgs theory it is also possible to
define a monopole charge. The monopole charge is indeed conserved (nice
detailed reviews are, for instance, \cite{vilenkin} \cite{ZJ1} \cite{west})
but, in order to prove this fact, one needs to use the \textit{boundary
conditions and the Bianchi identities} (instead of the equations of motion).
A similar situation occurs in scalar field theories in two dimensions which
admit kinks solutions: the kink number can be defined and one can prove that
the kink number is conserved using just the boundary conditions but without
using the equations of motion and the symmetries of the corresponding
action. For these reasons, the term "topological charge" is commonly used in
these cases. Furthermore, in the cases in which the gauge theory admits a
supersymmetric version, topological charges play an important role since
they enter the so called BPS bound (see, for instance \cite{west}) which
represents a (necessary) condition in order for classical solutions to
preserve a fraction of the supersymmetries of the theory. Roughly speaking,
the BPS bound can be written as%
\begin{equation}
M\geq \left\vert Q_{t}\right\vert  \label{bpsbound}
\end{equation}%
where $M$ is the mass of the classical solution and $Q_{t}$ is the
corresponding topological charge. When the bound is saturated one can expect
that the classical solution preserves some fraction of the supersymmetries.
As it is well known, BPS-like bounds play a very important role in the
stability analysis of classical backgrounds: even in situations in which no
supersymmetric argument is available, a BPS-like bound can help in proving
the stability of the corresponding solution.

In the case of Lovelock gravities, there is a close analogy between the role
of torsion and the non-Abelian magnetic field of a Yang-Mills theory which
suggests a natural ansatz (first proposed in \cite{canfo}) to find exact
solutions with non-trivial torsion. Thus, inspired by the case of
Yang-Mills-Higgs theories in which the BPS bound often implies some linear
relations among the electric and magnetic part of the field strength, the
idea is to consider a torsion which is a linear combination of the
components of the curvature which appears in the equations of motion (see
Eq. (\ref{ansatzt1})).\ This idea allowed to construct the first exact
vacuum solutions with non-trivial torsion in Lovelock gravities \cite{CGW} 
\cite{CGT} \cite{CG}. In particular, it has been shown in \cite{CGW} that
using the proposed ansatz for the torsion it is possible to construct a
class of black hole solutions in five dimensional Chern-Simons supergravity
with a ground state which preserves half of the supersymmetries (something
which, without torsion is known to be impossible). Therefore, being the
torsion necessary in order to achieve a half-BPS black-hole solution, it is
quite natural to expect that it should be possible to construct with torsion
a topological charge (namely, a charge which is constructed using the
Bianchi identities but without using the equations of motion or the
symmetries of the solution) which enters some sort of BPS bound as in Eq. (%
\ref{bpsbound}).

In order to provide one with a more concrete argument supporting the fact
that torsion could lead to the definition of a topological charge, one can
proceed as follows. We will construct for the special class of metric in Eq.
(\ref{8dcase}) a six-form $Q_{\Omega }$ such that $DQ_{\Omega }$ vanishes by
virtue of only the Bianchi identities.

It is useful to recall here the Bianchi identities corresponding to a
generic $N-$dimensional background metric with torsion (whose curvature
two-form and torsion two form read $R^{AB}$ and $T^{C}$ with $A,B,C=1,...,N$%
):%
\begin{equation}
DT^{C}=R_{CB}e^{B}\ ,\ \ \ \ DR^{AB}=0\ ,  \label{biantor}
\end{equation}%
where $e^{B}$ is the vielbein and $D$ is the covariant derivative$.$

Let us consider the two following three-forms 
\begin{equation}
\Omega _{(1)}=H(r)T^{a}e_{a}\ ,\ \ \ \Omega _{(2)}=H(r)T^{i}e_{i}\ ,
\label{proposalcharge}
\end{equation}%
where $\tilde{e}_{i}$ and $\tilde{e}_{a}$ are the intrinsic vielbeins of \ $%
M_{3}$ and $N_{3}$ and%
\[
H(r)=\frac{1}{r^{2}}\ . 
\]%
Using both the Bianchi identities in the presence of torsion and the
explicit expressions of the curvature in the presence of non-vanishing
torsion in Eqs. (\ref{modifiedcurvature}) and (\ref{modifiedcurvature2}) (%
\textit{but without using the equations of motion}), one easily gets that
the six-form $Q_{\Omega }$, defined as 
\begin{equation}
Q_{\Omega }=\Omega _{(1)}\wedge \Omega _{(2)}\ \ ,  \label{propcharge0}
\end{equation}%
is closed 
\[
DQ_{\Omega }=0\ . 
\]

Indeed, this property of the six-form $Q_{\Omega }$ is similar to the
property which is needed to construct a conserved charge in Lovelock
theories in the Komar approach \cite{kastor}. The usual strategy is to
evaluate the integral of a closed form on a spacelike slice of the
space-time in such a way to get two contributions, one coming from the
asymptotic region and another coming from the horizon. Then, using the fact
that the form is closed, one could conclude that the two contributions are
equal obtaining as a by-product the conserved charge as well. Interestingly
enough, using the expressions in Eq. (\ref{propcharge0}) one would obtain a
charge which is proportional to the product $\delta _{(1)}\delta _{(2)}$
where the integration constants $\delta _{(1)}$ and $\delta _{(2)}$
correspond to the presence of non-trivial torsion fluxes along both $N_{3}$\
and $M_{3}$. It is worth noting that the present proposal in Eq. (\ref%
{propcharge0}) would give a vanishing topological charge if only one of the
constants ($\delta _{(1)}$ or $\delta _{(2)}$) would be different from zero.
In order to go through with this method for the definition of the
topological charge associated with torsion, one should construct suitable
forms associated with torsion similar to the ones in Eq. (\ref{propcharge0})
which are closed due to \textit{only} the Bianchi identities for a generic
vacuum solution with torsion of Lovelock theories and not just for the class
of metrics in Eq. (\ref{8dcase}). Because of the scarcity of exact solutions
with torsion in Lovelock theories, this task seems to be quite difficult but
it certainly deserves further investigation.

A natural question (which is also closely related to the nature of the would
be topological charge which has been discussed above) is: what kind of
physical effects are expected in a background in which the torsion has the
form in Eq. (\ref{ansatzt1})? Indeed, it can be easily seen that geodesics,
in a background metric in Eq. (\ref{8dcase}) with a totally anti-symmetric
torsion as described in Eq. (\ref{ansatzt1}), are not affected by the
presence of torsion. On the other hand, the dynamics of spinors is affected
by the presence of torsion since the contorsion enters directly into the
Dirac equations. Thus, as it has been pointed out in \cite{CG}, in a
background with torsion the Dirac fields are, in a sense, "polarized" by the
torsion itself. For this reason, if it would be possible to define in
general a topological charge associated with (the totally antisymmetric part
of) torsion in Lovelock gravities then the nature of this charge would be
more like a spin than as a usual angular momentum.

\section{D=10 base manifold $M_{4}\times N_{4}$}

In order to emphasize the peculiarity of the base manifold $M_3\times N_3$
in the eight dimensional black hole solution, it is worth to analyze here
its most natural generalization to ten dimensions. In particular we will
show that in the ten dimensional case the entropy is not identically zero.
In this case the base manifold is the product of two four dimensional
constant curvature manifolds with curvature $\gamma $ and $\eta $. Once
again, it is convenient to normalize the mass parameter $\mu $ to one: $\mu
=1$. Then, the equations of motion give 
\begin{equation}
3(\gamma +1)^{2}(\eta +1)^{2}+24(1+\gamma )(1+\eta )+8=0  \label{10deq}
\end{equation}

An interesting feature of this case is that Eq. (\ref{10deq}) is quadratic
in $\eta $ and $\gamma $. Therefore, for a fixed $\eta $, there are two
possible values $\gamma _{\pm }$ of $\gamma $: 
\[
\gamma _{\pm }=-1-\frac{12(1+\eta )\mp \left\vert 1+\eta \right\vert \sqrt{%
120}}{3(\eta +1)^{2}} \label{paja} 
\]
so the solution admits two branches. The fact that solutions of Lovelock
gravity generally admit several branches is well known \cite{boulware-deser}%
and due to the fact that the theory is polynomial in the curvature. In this
context naturally arises the question if the two branches behave differently
from the thermodynamical point of view.\newline
The Wald entropy can be easily computed as function of $\eta$ and $\gamma$
and reads 
\begin{equation}
S=\kappa r_+ ^8c_1 Vol(\Sigma)\left[ 12(\gamma^2 \eta + \eta^2 \gamma) + 6
(\gamma^2 +\eta^2 +12\gamma \eta) +60(\gamma +\eta) +70 \right]
\end{equation}
In order to evaluate the entropy for the two branches it is necessary to
insert the corresponding expressions in Eq. (39). It can be seen that for
both branches the behavior is qualitatively identical and especially the
entropy does not change sign in both branches. Depending on the sign of the
overall coupling constant $\kappa $, the entropy can be negative definite or
positive definite. In the four-dimensional case, the overall factor $\kappa $
has to be chosen to be positive since, when $D=4$, a negative $\kappa $
would imply that the graviton is a ghost. On the other hand, in dimensions
higher than four, the BI Lovelock theory is degenerate when expended around
the maximally symmetric vacuum so that it is difficult to define a
meaningful graviton propagator so, when $D>4$, one cannot use the same
argument as in $D=4$ to fix the sign of $\kappa $. Indeed, as it has been
first noticed in \cite{CGTW}, it is even possible to take advantage of the
freedom to choose the overall sign in Lovelock gravities to provide one with
natural mechanism for spontaneous compactification. Therefore, in the
present case, it is possible to choose the constant $\kappa $ to be negative
in order obtain a positive definite entropy. The possibility of a negative
overall factor for a gravitational action, in order to get a positive
entropy, was argued also in \cite{cai2} in the context of the 3-D Lifshitz
black hole \cite{eloy}.

\section{Discussion and conclusions}

In the present paper, a new class of black hole solutions has been
constructed in even dimensional Lovelock Born-Infeld theory. These solutions
are interesting since, in some respects, they are closer to black hole
solutions of an odd dimensional Lovelock Chern-Simons theory than to the
more usual black hole solutions in even dimensions: this hybrid behavior
arises when non-Einstein base manifolds are considered. The thermodynamical
features of these solutions have been analyzed using Wald entropy. It turns
out that depending on the geometry of the base manifold the corresponding
entropy can change sign and even be zero with, at the same time, finite
temperature. Similar examples were already known in the literature on higher
order gravity theories. However, in such examples the origin of this
behavior lies in the choice of the coupling constants. In the present case
this peculiar behavior is due to the non-Einstein nature of the base
manifold. Indeed, when the base manifold is the product of different
constant curvature spaces, a new curvature scale appears which plays the
role of an integration constant. The sign of the entropy then depends on the
value of the integration constant. In this way the request of
thermodynamical stability constrains the geometry of the base manifold. It
has been also shown that in eight dimensions when the base manifold has the
form $M_{3}\times N_{3}$ (which leads to a vanishing entropy) a
non-vanishing torsion flux can appear. Eventually, the possible role of
torsion in the definition of a topological charge in Lovelock gravity has
been briefly discussed.

\section*{Acknowledgements}

{\small {The authors want to give a very warm thank for his illuminating
comments and suggestions to Julio Oliva, Sourya Ray and Ricardo Troncoso.
This work was supported by Fondecyt grant 11080056 and by UACh-DID grant
S-2009-57. The Centro de Estudios Cient\'{\i}ficos (CECS) is funded by the
Chilean Government through the Millennium Science Initiative, the Centers of
Excellence Base Financing Program of Conicyt and Conicyt grant "Southern
Theoretical Physics Laboratory" ACT-91. The work of F. C. has been partially
supported by PROYECTO INSERCI\'{O}N CONICYT 79090034 and A. S. I. (Agenzia
Spaziale Italiana). CECS is also supported by a group of private companies
which at present includes Antofagasta Minerals, Arauco, Empresas CMPC,
Indura, Naviera Ultragas and Telef\'{o}nica del Sur. 
}}

{\small \bigskip\ }

\end{document}